%% file: main.tex
\documentclass[runningheads]{llncs}

\usepackage{
  alphalph,
  amsmath,
  amssymb,
  amsfonts,
  booktabs,
  comment,
  colortbl,
  ifthen,
  listings,
  nicefrac,
  makecell,
  mathtools,
  microtype,
  multirow,
  paralist,
  rotating,
  setspace,
  semantic,
  stfloats,
  stmaryrd,
  subcaption,
  tabularx,
  times,
  wrapfig,
  xargs,
  xspace
}

\usepackage[inline]{enumitem}

\usepackage[
  pdftex,
  colorlinks = true,
  bookmarks = false,
  linkcolor = blue,
  citecolor = blue,
  urlcolor = blue,
  bookmarks,
  breaklinks = true
]{hyperref}

\usepackage[
  scale = .8
]{noto-mono}
\usepackage{cleveref}

\crefname{equation}{Eq.}{equations}
\Crefname{equation}{Equation}{Equations}
\crefname{proposition}{Prop.}{propositions}
\Crefname{proposition}{Proposition}{Propositions}
\crefname{table}{Tbl.}{tables}
\Crefname{table}{Table}{Tables}
\crefname{definition}{Def.}{definitions}
\Crefname{definition}{Definition}{Definitions}
\crefname{theorem}{Thm.}{theorems}
\Crefname{theorem}{Theorem}{Theorems}
\crefname{figure}{Fig.}{figures}
\Crefname{figure}{Figure}{Figures}
\crefname{page}{p.}{pages}
\Crefname{page}{Page}{Pages}
\crefname{section}{Sect.}{sections}
\Crefname{section}{Section}{Sections}

\crefname{lstlisting}{listing}{listings}
\Crefname{lstlisting}{Listing}{Listings}

\usepackage[textsize=scriptsize]{todonotes}

\input{macros.tex}
\title{Privacy with Good Taste}
\subtitle{A Case Study in Quantifying Privacy Risks in Genetic Scores\thanks{Work partially supported by the
    Danish Villum Foundation through Villum Experiment project
    No.~00023028 and New Zealand Ministry of Business, Innovation and
    Employment -- H\:\!\={\i}\:\!kina Whakatutuki through Smart Ideas
    project No.~UNIT1902.}}
\titlerunning{Privacy with Good Taste}

\author{Ra\'ul Pardo\inst{1} \and
Willard Rafnsson\inst{1} \and
Gregor Steinhorn\inst{2} \and
Denis Lavrov\inst{2} \and
Thomas Lumley\inst{3} \and
Christian W.~Probst\inst{2} \and
Ilze Ziedins\inst{3} \and
Andrzej W\k{a}sowski\inst{1}}

\authorrunning{R.~Pardo et al.}

\institute{%
  IT University of Copenhagen, Copenhagen, Denmark \and
    Unitec Institute of Technology, Auckland, New Zealand \and
    University of Auckland, Auckland, New Zealand
}

\usepackage[T1]{fontenc}
\begin{document}

\maketitle

\input{abstract.tex}

\input{introduction}

\input{background.tex}
\input{problem-statement}

\input{modelling}
\input{results}
\input{related}
\input{conclusion}

\bibliographystyle{splncs04}
\bibliography{privug}

\end{document}

%% file: macros.tex
\newcommand{\Dist}{\ensuremath{\mathcal{D}}}

\newcommand{\Int}{\mathbb{Z}}

\newcommand{\Normal}{\ensuremath{\mathcal{N}}}
\newcommand{\Uniform}{\ensuremath{\mathcal{U}}}

\newcommand{\Categorical}{\ensuremath{\mathcal{C}\textit{\hspace*{-0.4mm}at}}}

\newcommand{\Reals}{\ensuremath{\mathbb{R}}}
\newcommand{\Nats}{\ensuremath{\mathbb{N}}}

\newcommandx{\con}[4][2=,3=,4=]{\ensuremath{%
\textsf{#1}%
\ifthenelse{\equal{#2}{}}{}{\,#2}%
\ifthenelse{\equal{#3}{}}{}{\,#3}%
\ifthenelse{\equal{#4}{}}{}{\,#4}%
}}

\newcommand{\fun}[2]{\ensuremath{#1(#2)}}

\newcommand{\Dst}[1]{\ensuremath{\fun{\Dist{}}{#1}}}

\newcommand{\eg}{\textit{e.g.}}
\newcommand{\etal}{\textit{et al.}\xspace}

\newcommand{\privug}{{\sc Privug}\xspace}

\newcommand{\inputletter}{I}
\newcommand{\outputletter}{O}
\newcommand{\inputs}{\ensuremath{\mathbb{\inputletter}}}
\newcommand{\outputs}{\ensuremath{\mathbb{\outputletter}}}
\newcommand{\rvinput}{\ensuremath{\mathit{\inputletter}}}
\newcommand{\rvoutput}{\ensuremath{\mathit{\outputletter}}}

\newcommand{\haplotypeletter}{H}
\newcommand{\ethnicityletter}{E}

\newcommand{\sethaplotypes}{\ensuremath{\mathbb{\haplotypeletter}}}
\newcommand{\rvhaplotype}{\ensuremath{\mathit{\haplotypeletter}}}

\newcommand{\setethnicities}{\ensuremath{\mathbb{\ethnicityletter}}}
\newcommand{\rvethnicity}{\ensuremath{\mathit{\ethnicityletter}}}

\newcommand{\rthirtyeigth}{\ensuremath{\mathit{r38}}}
\newcommand{\rsixteen}{\ensuremath{\mathit{r16}}}
\newcommand{\gs}{\textit{gs}}
\newcommand{\gt}{\textit{gt}}
\newcommand{\Lgs}{\textit{L}_\gs}
\newcommand{\NL}{\textit{NL}}
\newcommand{\NLgs}{\NL_\gs}

\newcommand{\Ph}{\textit{Ph}}

\newcommand{\nt}{\textit{nt}}
\newcommand{\Phnt}{\ensuremath{\Ph_\nt}} 

\makeatletter
\newcommand\footnoteref[1]{\protected@xdef\@thefnmark{\ref{#1}}\@footnotemark}
\makeatother

\renewcommand{\subparagraph}[1]{\vspace{6.5pt plus 2.1pt minus 1pt}{\noindent\sffamily\itshape #1.}}

\newcommandx{\labelS}[2][1=7.8mm]{%
  \phantomsubcaption\label{#2}
  \llap{%
    \scriptsize%
    \raisebox{#1}[0mm][0mm]{%
      (\subref*{#2})                
    }\hspace{.2mm}%
  }}%

%% file: abstract.tex
\begin{abstract}
Analysis of genetic data opens up many opportunities for medical and scientific advances. The use of phenotypic information and polygenic risk scores to analyze genetic data is widespread. Most work on genetic privacy focuses on basic genetic data such as SNP values and specific genotypes. In this paper, we introduce a novel methodology to quantify and prevent privacy risks by focusing on polygenic scores and phenotypic information. Our methodology is based on the tool-supported privacy risk analysis method Privug. We demonstrate the use of Privug to assess privacy risks posed by disclosing a polygenic trait score for bitter taste receptors, encoded by TAS2R38 and TAS2R16, to a person's privacy in regards to their ethnicity. We provide an extensive privacy risks analysis of different programs for genetic data disclosure: taster phenotype, tasting polygenic score, and a polygenic score distorted with noise. Finally, we discuss the privacy/utility trade-offs of the polygenic score.

\end{abstract}





%% file: introduction.tex
\section{Introduction}%
\label{sec:intro}

Genetics strongly influence \emph{phenotypes}, the observable traits of humans and other species.  Since the successful sequencing of a human genome in 2003, many attempts have been made to develop new methods utilizing this vast information.  Research focuses on understanding the association between phenotypic and genetic information (see, \textit{e.g.}, \cite{2016_Risso,SORANZO20051257} on taste reception genes).  \emph{Polygenic risk scores} are developed to summarize the effect of genes on phenotype, especially in medical applications \cite{2018_polygenetic_risk_scores}. They are typically defined as a weighted sum on genetic data related to a single phenotype trait.  Unfortunately, the use of a genotype in a polygenic score could disclose information about other conditions it is associated with. For example, the Apolipoprotein E (ApoE) gene shows both strong correlation with cardiovascular disease risk and Alzheimer's disease risk~\cite{2020_Lumsden}.
\looseness = -1

Researchers have demonstrated privacy risks associated with genetic data\,\cite{acm_survey_genomic_privacy}.  For instance, an individual's genomic data can be used to find out predisposition to disease, \textit{e.g.}, using the phenotypic information or polygenic scores mentioned above.
A person's genome is based on their ancestry, with the addition of any mutations acquired by that person or their ancestors~\cite{2013_gymrek}.
%
As a consequence, disclosing genetic data poses privacy risks, not only for its owner, but also her relatives and ancestors~\cite{humbertCCS13}.
On a population level (not necessarily for individuals), knowledge about an individual's ancestry allows to make predictions about their ethnicity.
The distribution of genotypes for a population is based on ancestry, therefore genetic data correlates to the ethnicity of individuals (\textit{e.g.},~\cite{2016_Risso}).
This poses a privacy risk for individuals who may be subject to discrimination.
\looseness = -1

Most privacy risk analyses and anonymization mechanisms in genetics focus on basic genetic data---such as SNP values or specific genotypes~\cite{acm_survey_genomic_privacy}.
These approaches have been proven to be very effective in anonymizing and quantifying different kinds of privacy risks such as reidentification, kin privacy, or health care privacy~\cite{2015_cai,2020_Gursoy_and_Gerstein,2018_harmanci,humbertEuroSP2022}.
See~\cref{sec:related} for a detailed discussion of related works.

In this paper, we propose to quantify and prevent privacy risks by focusing on polygenic scores and phenotypic information.
To the best of our knowledge, this is the first work to explore this viewpoint to tackle genetic privacy.
Our work does not aim to replace existing methods, but to complement them through this new lens.
This work is motivated by the observation that genetic data is often disclosed in terms of phenotypic information and polygenic scores.
So it is directly applicable to the way geneticists process and disclose information.
We build on top of the privacy risk analysis method \privug~\cite{privug}.
Given a disclosure program (\textit{e.g.}, the program to compute a polygenic score), a probabilistic model of attacker knowledge and an output of the program, \privug\ computes the attacker posterior knowledge that can be used to assess privacy risks.
We demonstrate the use of \privug\ to assess the privacy risks posed by disclosing a polygenic trait score for the TAS2R38 and TAS2R16 taste receptor genes.
We quantify the risks for a person's privacy in regards to their ancestry and thereby derived their likely ethnicity.
The data and programs in this case study are selected to enhance readability and to serve as a template to apply our methodology.
The methodology we present can be applied to phenotypes and polygenic scores working on any kind of sensitive genetic data.
%
%
In summary, our contributions are:
\begin{itemize}
\item A methodology to analyze privacy risks of phenotipic information and polygenic scores based on the \privug method.
\item A demonstration of the methodology on real case study based on the TAS2R38 and TAS2R16 taste receptor genes and their correlation with ethnicity.
\item An extensive privacy risks analysis of different programs for genetic data disclosure: taster phenotype, tasting polygenic score, and a polygenic score distorted with noise.
\item An analysis of the trade-off between privacy and utility of the polygenic score.
\end{itemize}
The data and source code of all experiments are available at: \url{https://github.com/itu-square/privug-genetic-privacy}.

%% file: background.tex
\section{Background}%
\label{sec:back}


\noindent
\textit{Taster Genes.}
The \emph{genotype} is the genetic description of an organism made up of the specific alleles of genes an individual has inherited. A \emph{phenotype} is an observable trait of an organism, in our case, tasting bitterness or sourness.  Several studies found correlations between TAS genotypes (a fragment of the entire genotype of humans) and the perception of chemical substances~\cite{2013_Behrens,2002_Bufe,wine_tasting,2016_Risso}.
TAS2R38 is predominantly responsible for detecting bitterness\,\cite{2013_Behrens,2016_Risso} and TAS2R16 is associated with detecting sourness\,\cite{2002_Bufe}. Together they define the taster phenotype explored in this study. A \emph{haplotype} is (a part of) a genotype containing chromosomes from one parent only.  In this paper,  we focus on the pairs of haplotypes that compose the genotypes of TAS2R38 and TAS2R16.  We do not consider more basic elements such as \emph{alleles}.  The haplotypes of TAS2R38 are PAV, AVI, AAV, AVV, PAI, PVI, AAI and PVV.  The haplotypes of TAS2R16 are HAP-CD, HAP-A, HAP-B.  Thus, a given individual has a pair of haplotypes for each TAS2R genotype.
\looseness = -1

\noindent
\textit{Data Privacy Analysis with \privug.}
\label{sec:back_privug}
\privug\ is a tool-supported method to explore information leakage properties \noindent
of data analytics programs\,\cite{privug}. \privug\ assumes that a program transforms an input dataset to an output, which is subsequently disclosed to a third party called an \emph{attacker}.  \privug\ does not require a dataset, but starts with a probabilistic model of the attacker's knowledge.  The model is analyzed together with the program to study the risks of inference of sensitive information.

Let \inputs, \outputs\ denote sets of inputs and outputs of a program.  Let \Dst{\inputs} be a distribution over a set, in this case the set of inputs.
We write $\rvinput \sim \Dst{\inputs}$ to denote a random variable over the set of inputs.  The \privug\ method is divided in the following five steps:

\emph{Step 1: Attacker's Prior Knowledge.}
We model what the attacker knows about the input before observing the output of the program as a belief distribution.  For a program that receives an integer (\(\inputs \triangleq \Int\)), this could be a distribution \(\Uniform(-10,10)\), a discrete uniform distribution on integers between -10 and 10, which models an attacker knowing \emph{only} that the input is between $-10$ and $10$ but not more.  We write $p(\rvinput)$ for the probability distribution associated with the random variable \rvinput\ representing the input to the program.

\emph{Step 2: Interpret the Program.}
We run the program not on a concrete input data set from\,\inputs, but on the belief distribution representing the attacker's knowledge about the input.  For example, the following program takes as input an integer and returns its value perturbed by a Laplacian distribution with mean 0 and scale 1:
\begin{lstlisting}
def program(x: int): return x + stats.laplace.rvs()
\end{lstlisting}
We transform this program into a probabilistic one taking a distribution over inputs and run it on the attacker's knowledge distribution:
\begin{lstlisting}
def program(x: Dist(int)): return x + stats.laplace.rvs()
\end{lstlisting}
where \lstinline[language=Python]{Dist(int)} denotes a distribution over integers (\Dst{\Int}).  The attacker's knowledge together with the program define the joint distribution over inputs and outputs: \(p(\rvinput,\rvoutput)\).
\looseness = -1

\emph{Step 3: Observation.}
Optionally, we can assume that the output of the program, or some information about it, has been disclosed to the attacker (otherwise we reason about all possible input data sets).  For instance, assume that the attacker learned that the output of the program was greater than 7.  Adding this observation amounts to conditioning the  joint distribution, \eg switching from \(p(\rvinput,\rvoutput)\) to \(p(\rvinput,\rvoutput \mid \rvoutput > 7)\).

\emph{Step 4: Posterior.}
We approximate the joint distribution using standard Markov Chain Monte Carlo (MCMC) methods.  In this paper, we use the PyMC3~\cite{pymc3} library.
\looseness = -1

\emph{Step 5: Posterior Analysis.}
We query the inferred distribution to study the posterior knowledge of the attacker.
To this end, we can query for probabilities and compute summary statistics of the distributions (mean, variance, etc.), and standard leakage measures such as entropy, KL-divergence, mutual information, and Bayes vulnerability\,\cite{qif}.
%


%% file: problem-statement.tex
\section{The Case Study}%
\label{sec:problem-statement}

\begin{figure*}[t!]

  \includegraphics[
    width = \textwidth,
    trim = 5mm 3mm 1mm 5mm,
    clip
  ]{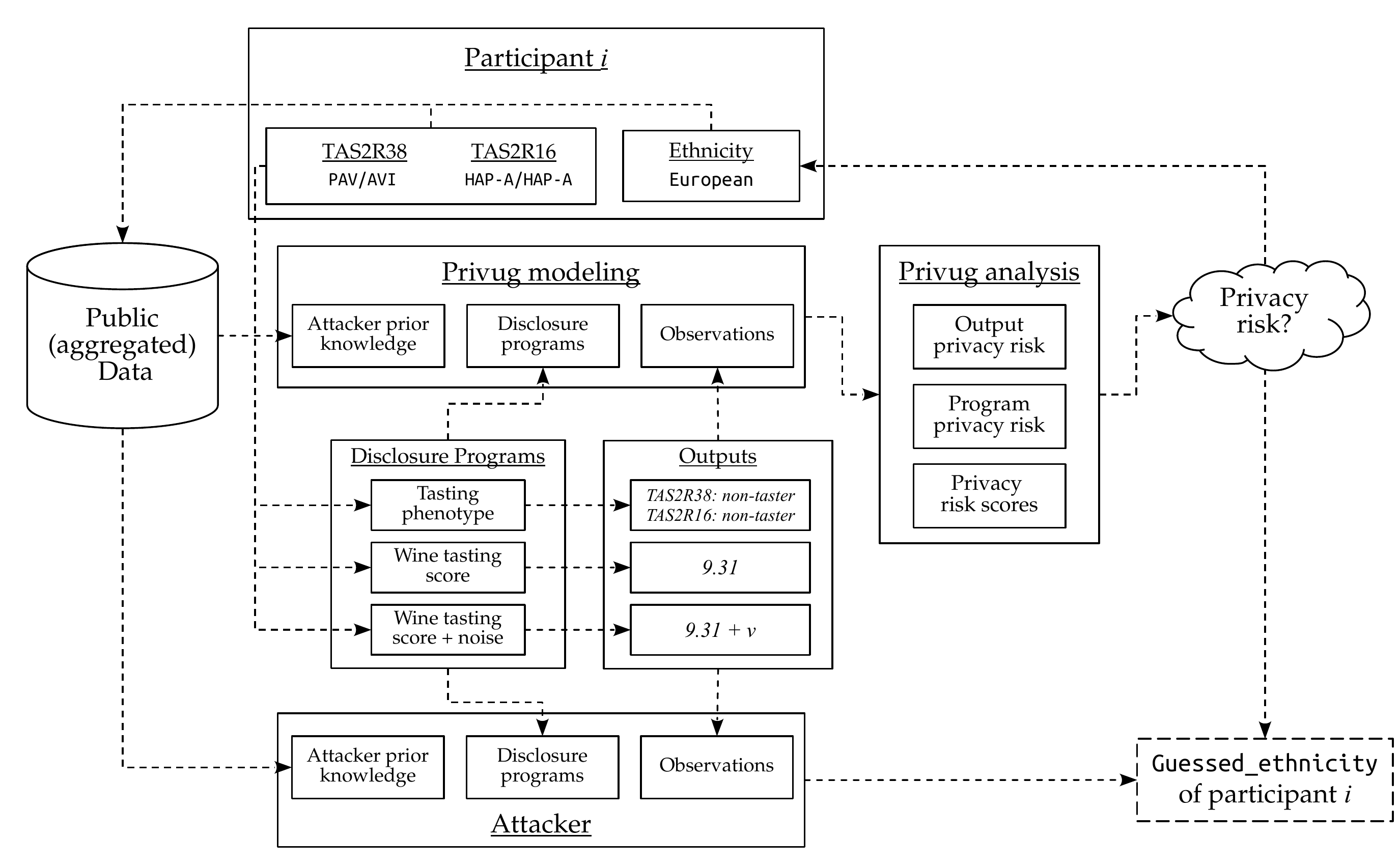}
  \caption{The case study overview.}%
  \label{fig:problem-overview}

  \vspace{-2mm} 

\end{figure*}


A data analyst wants to disclose data about the ability of study participants to taste wine.  Such data is commonly released~\cite{2013_Behrens,2002_Bufe,wine_tasting,2016_Risso}.  To compute the tasting information, the analyst uses the information about the taste receptor genes TAS2R38 and TAS2R16.
\Cref{fig:problem-overview} includes an example of data for a single participant in the box labeled \textit{Participant \textit{i}}, including haplotype pairs \texttt{PAV/AVI} for TAS2R38 and \texttt{HAP-A/HAP-A} for TAS2R16.  The analysts considers the following three options of disclosing the data.

\noindent
\textit{1.~Taster/Non-taster binary phenotype.}
This program labels participants as \emph{taster}, who can taste bitterness and sourness (having relevant haplotypes of TAS2R38 and TAS2R16), or \emph{non-taster}.  For Participant \textit{i} the output of this program is \emph{non-taster} on both accounts.
\looseness = -1

\noindent
\textit{2.~Wine tasting score / polygenic score.}
Combines TAS2R38 and TAS2R16 haplotype pairs to compute a genetic trait score. 
%
%
The polygenic score is based on biochemical tests, published in~\cite{2013_Behrens}, to determine the response of TAS2R38 haplotypes to bitter substances, and the presence TAS2R16 taster haplotypes.
The larger the score, the better the wine tasting abilities of the participant.  For Participant \textit{i} the program output is 9.31.

\noindent
\textit{3.~Wine tasting score with noise.}
In the spirit of differential privacy, this program adds noise to the output of the previous one with the goal of decreasing privacy risks.  For Participant\,\(i\) it outputs 9.31 plus a random perturbation $\nu$ drawn from a Normal distribution with mean 0 and standard deviation $\sigma$.  The value of $\sigma$ determines the amount of noise.
\looseness = -1

\noindent
The first two disclosure programs are standard methods to aggregate and share genetic data.  They are not designed with privacy protection in mind.  The last method attempts to enhance privacy by adding random noise to the wine tasting score.

In this case study, the ethnicity of participants is considered sensitive information.  Note that our programs do not use ethnicity as input.  Still, their output could be used to learn about ethnicity using a linking attack.  Genetic information is correlated with ethnicity, and in this case the attacker may conclude that Participant\,\(i\) has European ethnicity.
\looseness = -1

We consider an information-theoretical attacker that has access to:
\begin{enumerate*}[label={\roman*)}]

  \item publicly available aggregated data correlating TAS2R38, TAS2R16 and ethnicity~\cite{2013_Behrens,2002_Bufe,wine_tasting,2016_Risso},  in particular \cref{tab:tas2r};

  \item the source of the disclosure program, and

  \item the program output released by the data analyst.

\end{enumerate*}
This is depicted in \cref{fig:problem-overview} as lines connecting those elements to the attacker model at the bottom.  The goal of the attacker is to infer the ethnicity of a study participant.  There are no bounds on the computational resources available to the attacker.

Our objective is to apply the \privug\ method to reason about this case, to expose privacy risks involved in releasing genetic data, as well as to encourage geneticists to consider \privug\ (and similar tools) as an aid in decision making.

%% file: modelling.tex
\section{Modeling}%
\label{sec:modelling}


In the following, we use $\sethaplotypes$ to denote the set of TAS2R38 \emph{haplotypes}: PAV, AVI, AA, etc. (second row in \cref{tab:tas2r}).  We use \setethnicities\ to denote the set of \emph{ethnicities} (first column in \cref{tab:tas2r}). We use \(H\) and \(E\) to denote the corresponding random variables.  We consider an attacker who, \emph{a priori}, makes no assumptions about the ethnicity of the participant.  In other words, before observing the output of the program, the attacker considers all ethnicities in \cref{tab:tas2r} to be equally likely. For convenience, we map each ethnicity to an element in $\Nats_0$. The prior is uniform over the ethnicities, i.e.\ \(E \sim \Uniform(0,3)\).




For the taste receptor haplotypes we consider an attacker informed by publicly available population genetics studies~\cite{2016_Risso,SORANZO20051257} containing information about the correlation between ethnicity, TAS2R38, and TAS2R16.  Given an ethnicity $\rvethnicity$, we use $\rthirtyeigth_\rvethnicity$ and $\rsixteen_\rvethnicity$ to refer to vectors composed by columns 2-9 and columns 10-12 in row $\rvethnicity$ of in~\cref{tab:tas2r38}, respectively, so we have that \(\rvhaplotype^\rthirtyeigth \sim \Categorical(\rthirtyeigth_\rvethnicity)\) and \(\rvhaplotype^\rsixteen \sim \Categorical(\rsixteen_\rvethnicity)\), where, for example, $\Categorical(\rthirtyeigth_\rvethnicity)$ is a categorical (discrete) distribution defined by vector \(\rthirtyeigth_\rvethnicity\).  Here again each haplotype value is mapped to an element in $\Nats_0$.

\begin{table*}[t!]
\setlength{\tabcolsep}{1.6pt}
\small
\begin{tabularx}{\linewidth}{
  X
  c c c c c c c c @{\hspace{15pt}} c c c
 }
\multicolumn{1}{c}{} & \multicolumn{8}{c}{\textbf{TAS2R38}} & \multicolumn{3}{c}{\textbf{TAS2R16}} \\[.5mm]
\multicolumn{1}{c}{} & PAV    & AVI    & AAV    & AVV    & PAI   & PVI   & AAI     & PVV    & HAP-CD   & HAP-A   & HAP-B \\
\midrule
African  & .5076 & .4270 & .0248 & .0032 & .0018 & .0007 & .0339 & .0010 &   .1511 & .8355  & .0133  \\
Asian    & .5076 & .3518 & .0061 & .0008 & .0000 & .0015 & .1322 & .0000 &   .0011 & .6309  & .3679  \\
European & .6451 & .3531 & .0000 & .0017 & .0000 & .0000 & .0000 & .0000 &   .0000 & .6810  & .3189  \\
American & .4566 & .4922 & .0356 & .0049 & .0032 & .0003 & .0055 & .0017 &   .0000 & .8105  & .1894  \\
\end{tabularx}

\medskip

\caption{TAS2R38 and TAS2R16 haplotype probability for each ethnicity, $p(H^{\rthirtyeigth}|E)$ and $p(H^\rsixteen|E)$, respectively \cite{2016_Risso,SORANZO20051257}}
\label{tab:tas2r38}
\label{tab:tas2r16}
\label{tab:tas2r}

\vspace{-2.5mm} 
%
\end{table*}

\begin{figure}[t!]

  \includegraphics[
    width = \textwidth
  ]{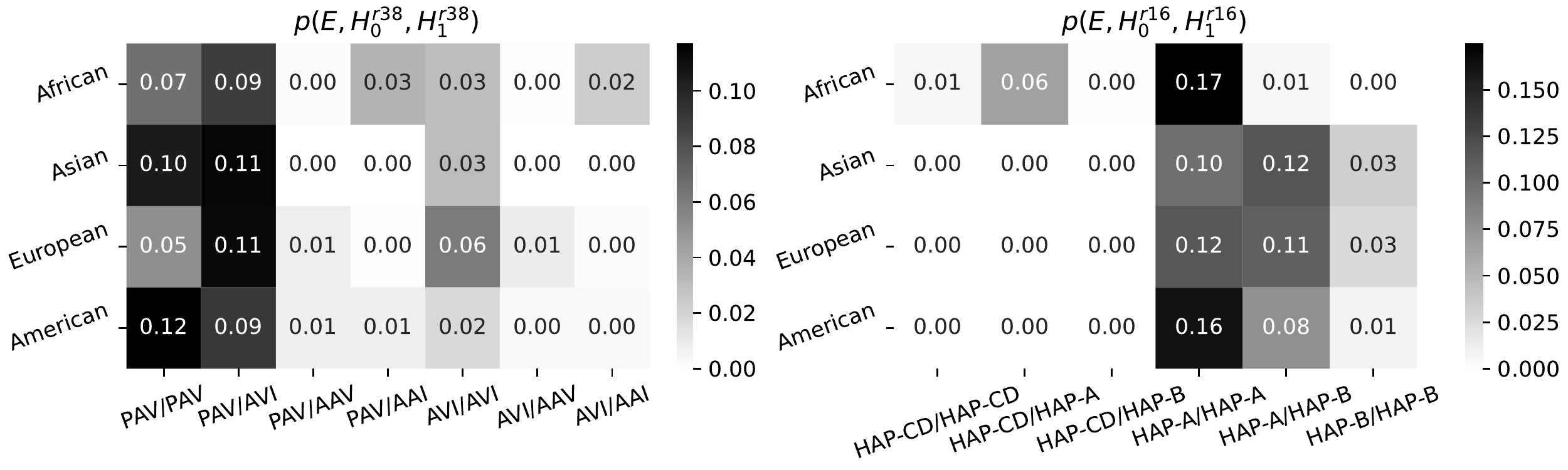}

  \caption{Priors on ethnicity, and haplotype pairs. Left: TAS2R38. Right: TAS2R16.}\label{fig:prior}

  \vspace{2.5mm} 

\end{figure}

\Cref{fig:prior} shows the joint distributions of ethnicity and haplotype pairs representing the beliefs of the attacker.
For instance, the top-left cell (left graph) shows that the probability of African ethnicity and the haplotype pair PAV/PAV is $0.07$. 
For TAS2R38, this prior assigns high probability to haplotype pairs PAV/PAV and PAV/AVI---as~\cite{2016_Risso} established that PAV and AVI are common haplotypes in all tested populations. 
For TAS2R16, the haplotype pairs HAP-A/HAP-A and HAP-A/HAP-B are most likely, due to the common occurrence of HAP-A~\cite{SORANZO20051257}. 

We now investigate and compare the privacy risks of the three disclosure programs from the previous section. All these work on pairs of haplotypes: we use $(H^\rthirtyeigth_0, H^\rthirtyeigth_1)$ to refer to each TAS2R38 haplotype and $(H^\rsixteen_0, H^\rsixteen_1)$ for TAS2R16.


\noindent
\textit{Taster/non-taster binary phenotype.}
\label{para:taster}
We consider a disclosure program that maps TAS2R38 and TAS2R16 to binary phenotypes: \emph{taster}, \emph{non-taster}.
For TAS2R38, the haplotype pair AVI/AVI corresponds to the non-taster phenotype, and the remaining haplotype pairs to taster.
For TAS2R16, the haplotype pair HAP-CD/HAP-CD corresponds to the non-taster phenotype, and the remaining haplotype pairs to taster:
\begin{equation*}
\Phnt^\rthirtyeigth \triangleq\hspace*{-1ex}\bigwedge_{i \in \{0,1\}}\hspace*{-1ex}H^\rthirtyeigth_i = \text{AVI}
\;\;\;\;\;\;\;\;\;\;\;
\Phnt^\rsixteen    \triangleq\hspace*{-1ex}\bigwedge_{i \in \{0,1\}}\hspace*{-1ex}H^\rsixteen_i = \text{HAP-CD}
\vspace*{-.5ex}
\end{equation*}
\Cref{code:taster-disclosure} shows their Python implementations.  Both programs take a haplotype pair and return a Boolean stating whether the pair corresponds to a non-taster.
\looseness = -1

\noindent
\textit{Wine tasting score / polygenic score.}
The polygenic score is a linear combination of genotype weights and haplotype weights:
$$
\Lgs \triangleq \alpha_{\rthirtyeigth}\cdot\gt_{\rthirtyeigth}(H^\rthirtyeigth_0,H^\rthirtyeigth_1) + \alpha_{\rsixteen}\cdot\gt_{\rsixteen}(H^\rsixteen_0,H^\rsixteen_1)
$$
The function $\gt_j \colon \sethaplotypes \times \sethaplotypes \to \Reals$ is the genotype weights; it assigns a score modeling the impact on tasting ability of a pair of haplotypes.
The coefficient $\alpha_j \in \Reals$ is the gene weight; it assigns a score modeling the influence of the gene on the tasting score.
We set the weight values based on the biochemical test~\cite{2013_Behrens}, for the response of TAS2R38 to bitter substances, and the presence TAS2R16 taster haplotypes.
%
%
%
\Cref{code:linear-gs} shows Python code for the polygenic score. 
The implementation rounds the value to 2 decimal points.
This is how polygenic scores are normally disclosed (and perceived by the attacker).


\begin{figure}[t!]
\begin{lstlisting}
def non_taster_TAS2R16(h1,h2): return h1==HAP-CD and h2==HAP-CD
def non_taster_TAS2R38(h1,h2): return h1==AVI and h2==AVI
\end{lstlisting}

\vspace{-2mm}

\caption{Taster disclosure program for TAS2R16 (line 1) and for TAS2R38 (line 2)}%
\label{code:taster-disclosure}

\begin{lstlisting}
def linear_gs(r38_h1,r38_h2,r16_h1,r16_h2):
  l_gs = $\alpha$_r38*gt_38(r38_h1,r38_h2) + $\alpha$_r16*gt_16(r16_h1,r16_h2)
  return round(l_gs,2)
\end{lstlisting}
\vspace{-2mm}
\caption{Wine tasting linear polygenetic score disclosure program}%
\label{code:linear-gs}

\begin{lstlisting}
def noisy_linear_gs(r38_h1,r38_h2,r16_h1,r16_h2):
  nl_gs = linear_gs(r38_h1,r38_h2,r16_h1,r16_h2) + np.random.normal(0,$\sigma$)
  return round(nl_gs,2)
\end{lstlisting}

\vspace{-2mm}

\caption{Wine tasting linear polygenetic score disclosure program with random noise}%
\label{code:noisy-gs}
\end{figure}

\noindent
\textit{Wine tasting score with random noise.}
\label{para:score-noise}
Here we consider a polygenic score aimed at reducing privacy risks.
We use a normal distribution with mean 0 and different values of standard deviation to generate random noise:
\begin{align*}
  \nu   \sim \Normal(0,\sigma) & & \NLgs \triangleq \Lgs + \nu
\end{align*}
\Cref{code:noisy-gs} shows a Python implementation.  The function \lstinline|np.random.normal(0,$\sigma$)| uses the NumPy\,\cite{numpy} library to sample from a normal distribution with mean 0 and standard deviation $\sigma$.  We do not fix the value of $\sigma$, to study the effect of increasing values in~\cref{sec:analysis}.
\looseness = -1



\section{Privacy Risk Metrics}
\label{sec:queries}
\label{sec:modeling-queries}


\paragraph{Output privacy risk.}
\label{sec:output-privacy}
We evaluate the privacy risk associated to disclosing a concrete output of the disclosure method.
To this end, we look at the posterior distribution of ethnicity given a concrete program output.
Let $O$ denote a random variable modeling the output of any of the programs in~\cref{sec:modelling}, we compute
$$
p(\ethnicityletter \mid O = v) \text{ for a concrete output } v \text{ in the domain of } O.
$$
If the probability for an ethnicity is high, then it means that the attacker can learn with high probability the ethnicity of the individual.
Output privacy is useful when a data analyst is trying to decide whether or not to disclose a program output.
For instance, in the taster phynotype program for TAS2R38, suppose that $p(\ethnicityletter=\text{African} \mid \Ph^\rthirtyeigth = \mathit{taster}) = 1$.
Now consider a data analyst that after running the program obtains \textit{taster}.
Then, releasing that output also discloses the individual's African ethnicity.

\noindent
\textit{Program privacy risk.}
\label{sec:program-privacy}
To evaluate the overall privacy risks of a program, we use a metric that accounts for the probability of each output, $p(O)$.
Note that output privacy measures risks disregarding how likely the output is.
Naturally, combining output privacy with the probability of the output yields the joint distribution of ethnicities and outputs,
$$
p(\ethnicityletter \mid O)p(O) = p(\ethnicityletter, O).
$$
Program privacy is useful for data analysts assessing risks before computing a concrete output.
High values indicate both a high risk of leaking the individual's ethnicity and that it is likely that the leak may occur.
Suppose that, for the taster TAS2R38 phynotype program, we have that $p(\ethnicityletter=\text{American}, \Ph^\rthirtyeigth = \mathit{taster}) = 0.8$.
That is, if the program outputs taster and the ethnicity of the individual is American with probability $0.8$, indepent of the input.
%
%
Intuitively, this program has high privacy risks for Americans.
Ideally, the program should distribute probability among ethnicities and outputs uniformly.

%

\noindent
\textit{Privacy risk scores.}
\label{sec:bayes-vulnerability}
These scores aim to summarize the output and program privacy risks into a single score (real value).
We use two privacy risk metrics to summarize privacy risks into a score: \emph{maximum output privacy} and \emph{Bayes vulnerability}~\cite{qif}.
%

\noindent
\textit{Maximum output privacy.}
This metric summarizes the results of output privacy risks.
It reports the maximum output privacy for all possible program outputs.
That is,
$$
\max_{e \in \ethnicityletter, o \in O}p(\ethnicityletter=e|O=o)
$$

Maximum output privacy is a \emph{pessimistic} upper bound on privacy risks, as it is pessimistic because it does not take into account the probability of the output.
A program may have large maximum output privacy for an output that is very unlikely.
Recall the example above where $p(\ethnicityletter=\text{African} \mid \Ph^\rthirtyeigth = \mathit{taster}) = 1$ and $P(\Ph^\rthirtyeigth = \mathit{taster}) = 0.01$.
Here the maximum output privacy equals 1.
Note that we do not need to explore other outputs; as 1 is the maximum output privacy risk.
%
%
This metric does not indicate what/how many outputs or ethnicities produce the maximum output privacy.
However, since maximum output privacy is an upper bound on privacy risks, a low value of output privacy does indicate low risks for all outputs.

\noindent
\textit{Bayes vulnerability.}
This metric summarizes program privacy risks.
Bayes vulnerability~\cite{qif} measures the \emph{expected probability of correctly guessing the ethnicity by observing the output of the program}.
Bayes vulnerability is defined as
$$
\mathit{V} = \sum_{o \in O} \max_{e \in \ethnicityletter} p(\ethnicityletter = e, O = o).
$$
A high value of Bayes vulnerability implies high privacy risks.
Bayes vulnerability does not indicate what ethnicity is at risk or what output causes the leak.
Bayes vulnerability is specially useful when comparing disclosure programs.
The joint distribution (program privacy risk) may consist of a large number of ethnicity/output pairs, making it tedious to compare among several programs.
Furthermore, Bayes vulnerability can be used as a first indicator of privacy risks.
In case Bayes vulnerability is high, then the joint distribution may be explored to find the ethnicities at high risk.


\section{Utility Metrics}
\label{sec:utility}


\noindent
\textit{Absolute difference.}
We consider the absolute difference of the wine tasting score (real output) and the wine tasting score with noise (distorted output), \textit{i.e.}, $|\Lgs - \NLgs|$.
A value of 0 indicates perfect utility, the larger the value the worse the utility.
Since our analysis estimates distributions $p(\Lgs)$ and $p(\NLgs)$, we actually analyze the distribution of the absolute difference, $p(|\Lgs - \NLgs|)$.

\noindent
\textit{Error bound probability.}
As for privacy risk metrics, now we define a score that summarizes utility.
Specifically, we consider the probability that the absolute difference is within a bound $\delta$, formally, $p(|\Lgs - \NLgs| < \delta)$.
The value $\delta$ defines the amount of error that the analyst considers acceptable.
For this paper, we (arbitrarily) set to study $\delta \in \{0.1, 0.5, 1\}$, but our analysis can be applied for any $\delta$.
High error bound probability indicates high utility, with 1 being perfect utility and 0 worst utility.


%% file: results.tex
\section{Analysis \& Results}
\label{sec:analysis}

In this section, we discuss:
\begin{inparaenum}[i)]
\item the quality of the inferred posterior distribution;
\item privacy risks of each disclosure program using the privacy risk metrics presented in~\cref{sec:modeling-queries}; and
\item the utility evaluation for the disclosure programs adding random noise.
\end{inparaenum}



\begin{figure*}[t!]
  \centering
  \includegraphics[trim=4 0 5 0, clip, width=.49\textwidth]{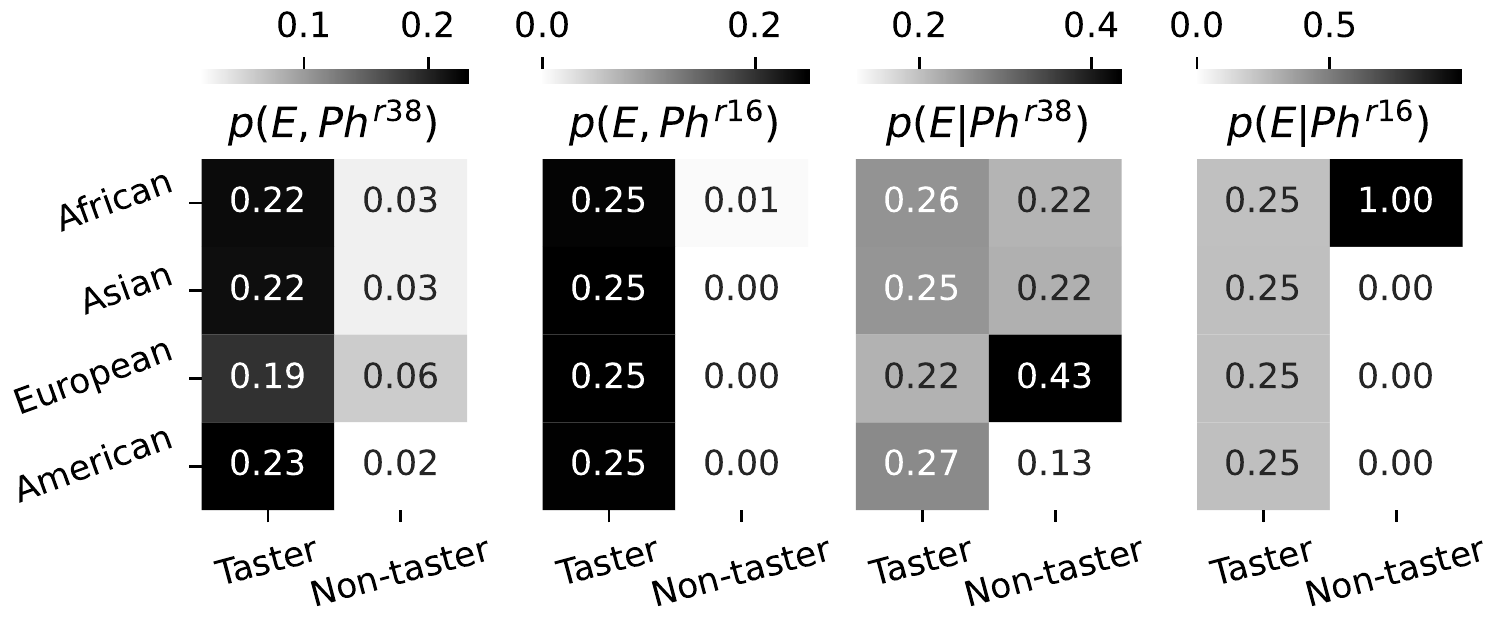}
  \includegraphics[trim=0 10 0 0, clip, width=.43\textwidth]{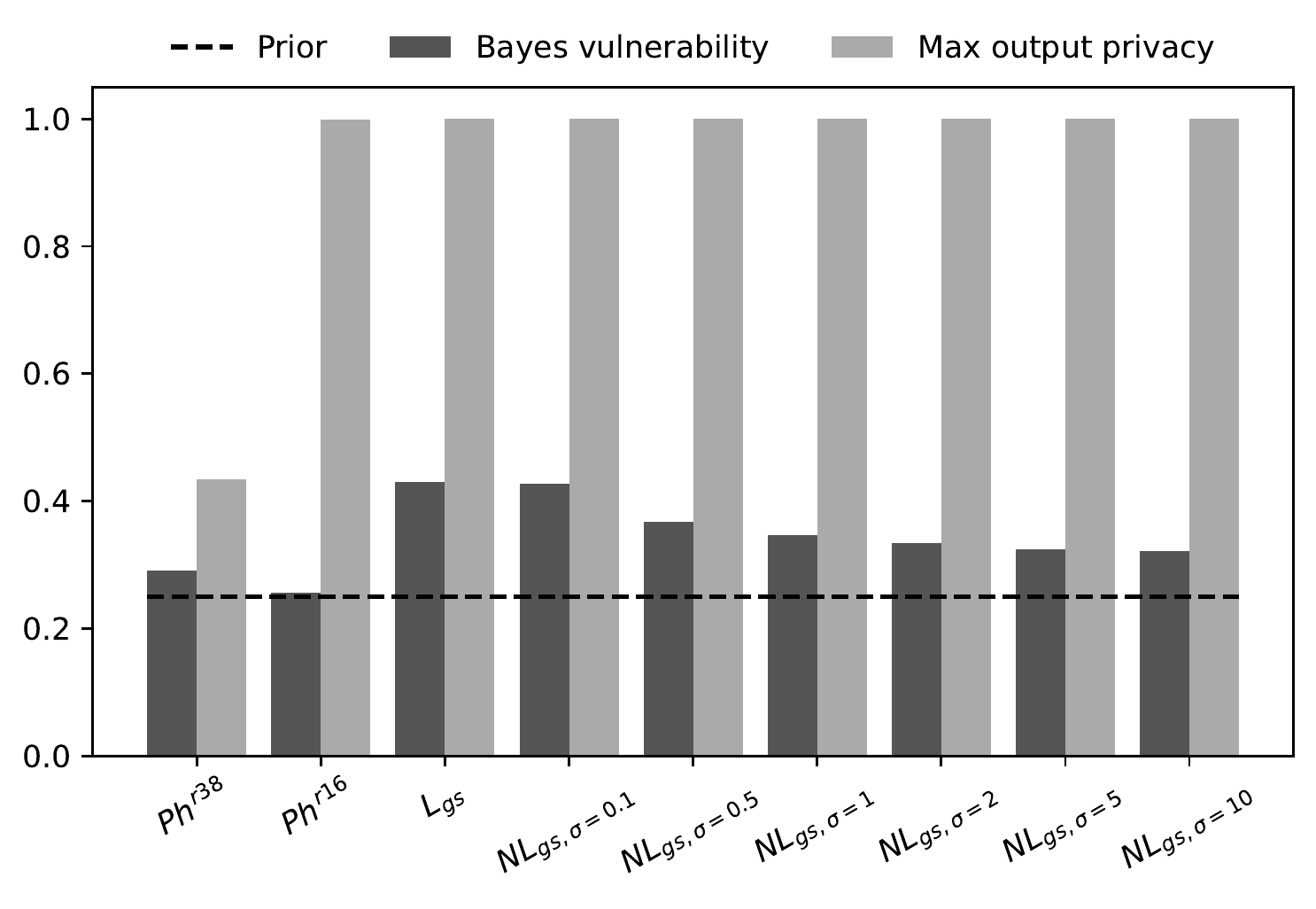}
  \vspace*{2mm}
  \includegraphics[width=\textwidth]{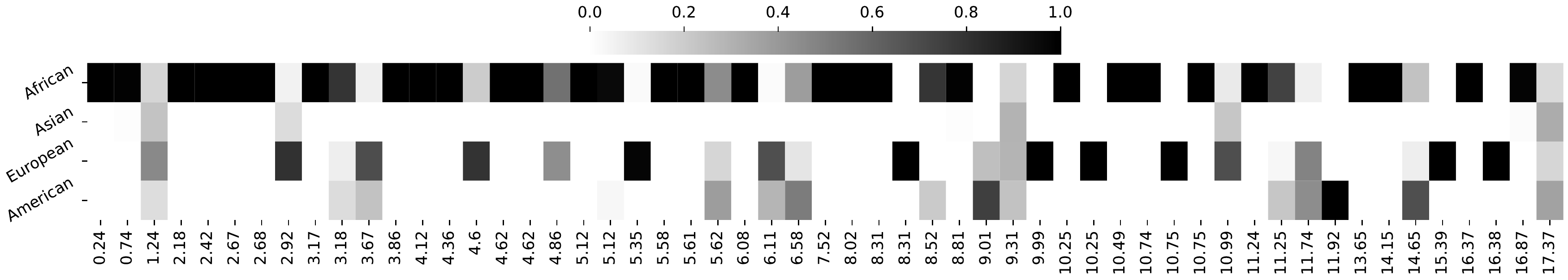}
  \vspace*{2mm}
  \includegraphics[width=\textwidth]{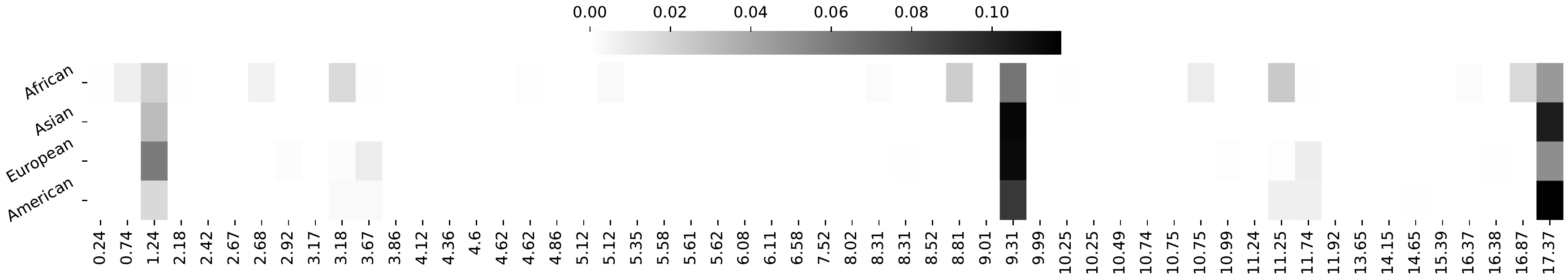}
  \caption{
    Top-left: Program privacy risks of binary phenotype program (first two histograms), and output privacy risk of binary phenotype program (3rd and 4th histograms).
    Top-right: Privacy risk scores results for all disclosure programs.
    Middle: Output privacy risk of polygenic scores.
    Bottom: Program privacy risk of polygenic score.}
  \label{fig:results-phenotype}
  \label{fig:results-linear-gs-conditional}
  \label{fig:results-linear-gs-joint}
  \label{fig:results-max-output-privacy-risk}
  \label{fig:results-bayes-vulnerability}
  \label{fig:results-bayes-vulnerability-and-max-output-risk}
\end{figure*}


\begin{figure}[t!]
  \centering
  \includegraphics[width=0.6\textwidth]{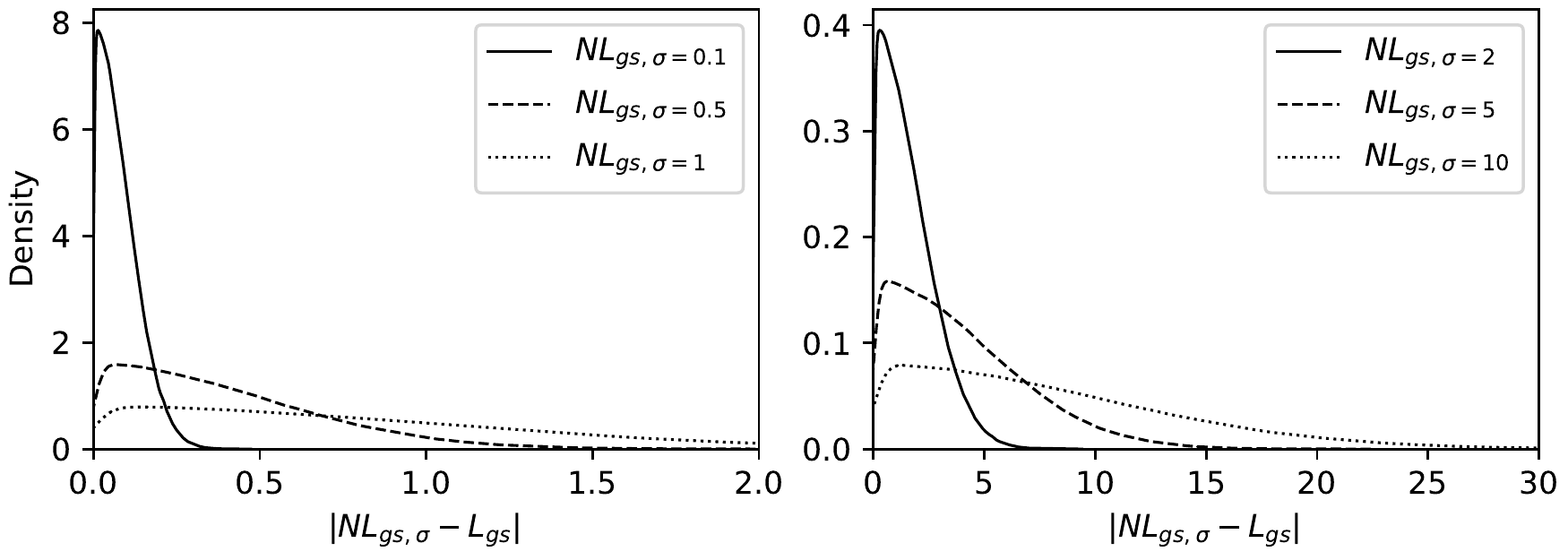}
  \includegraphics[width=.355\textwidth]{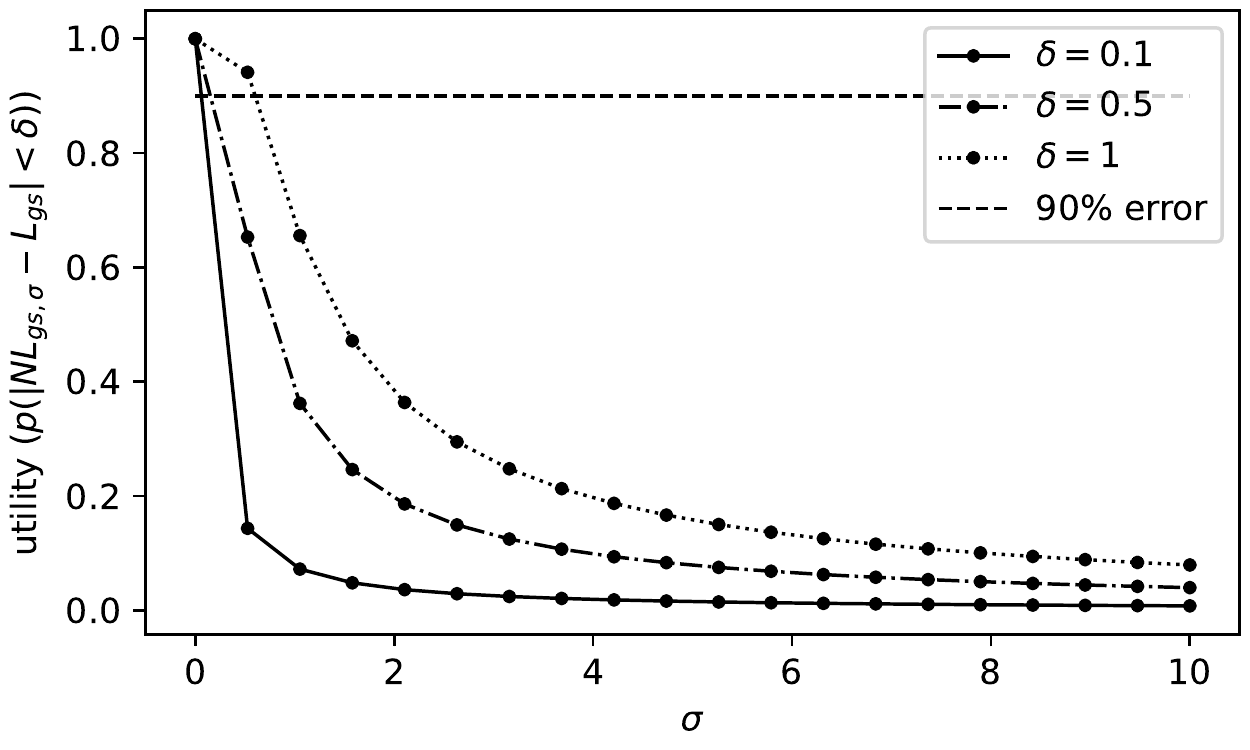}
  \caption{
    Utility results.
    Left, center: Distribution of absolute distance between noisy and real polygenic score. 
    Right: Error bound probability for different $\sigma$ and $\delta$. 
  }
  \label{fig:results-utility-metric}
  \label{fig:results-utility-sigma}
\end{figure}

\begin{figure*}[t!]
  \centering
  \includegraphics[trim=0 0 0 0,  clip, width=.36\textwidth]{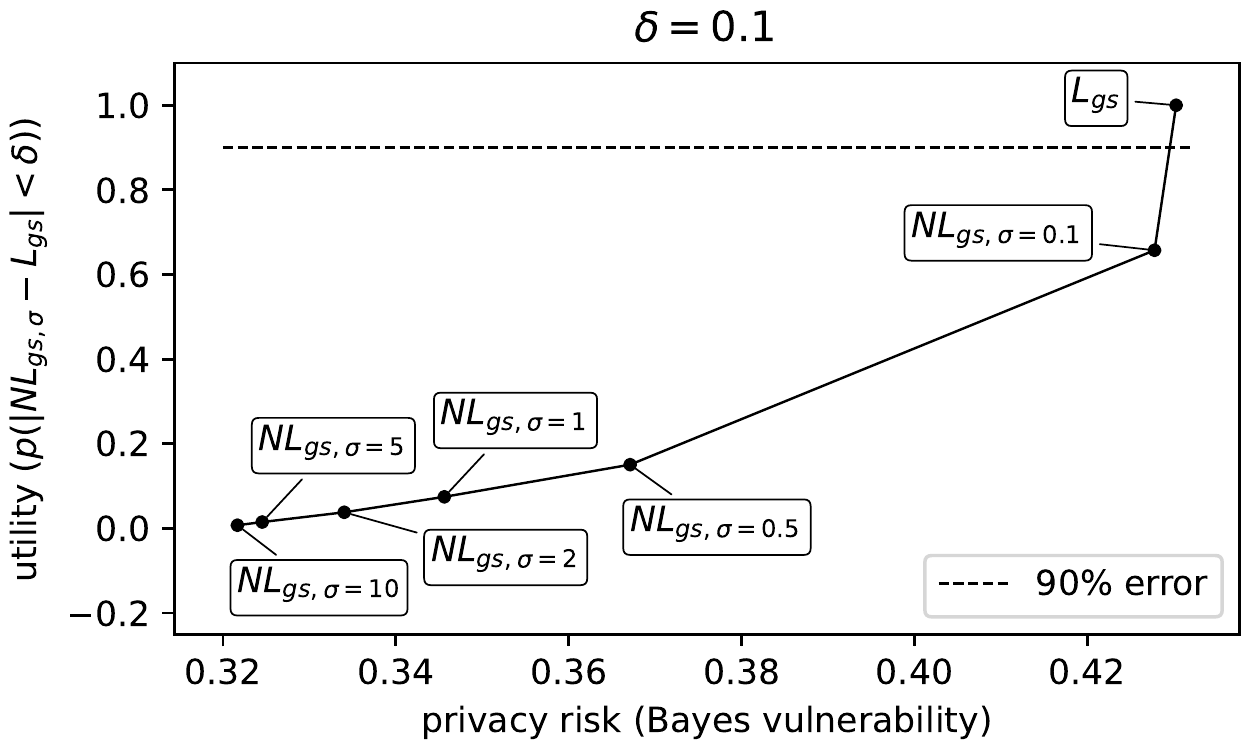}
  \includegraphics[trim=50 0 0 0, clip, width=.31\textwidth]{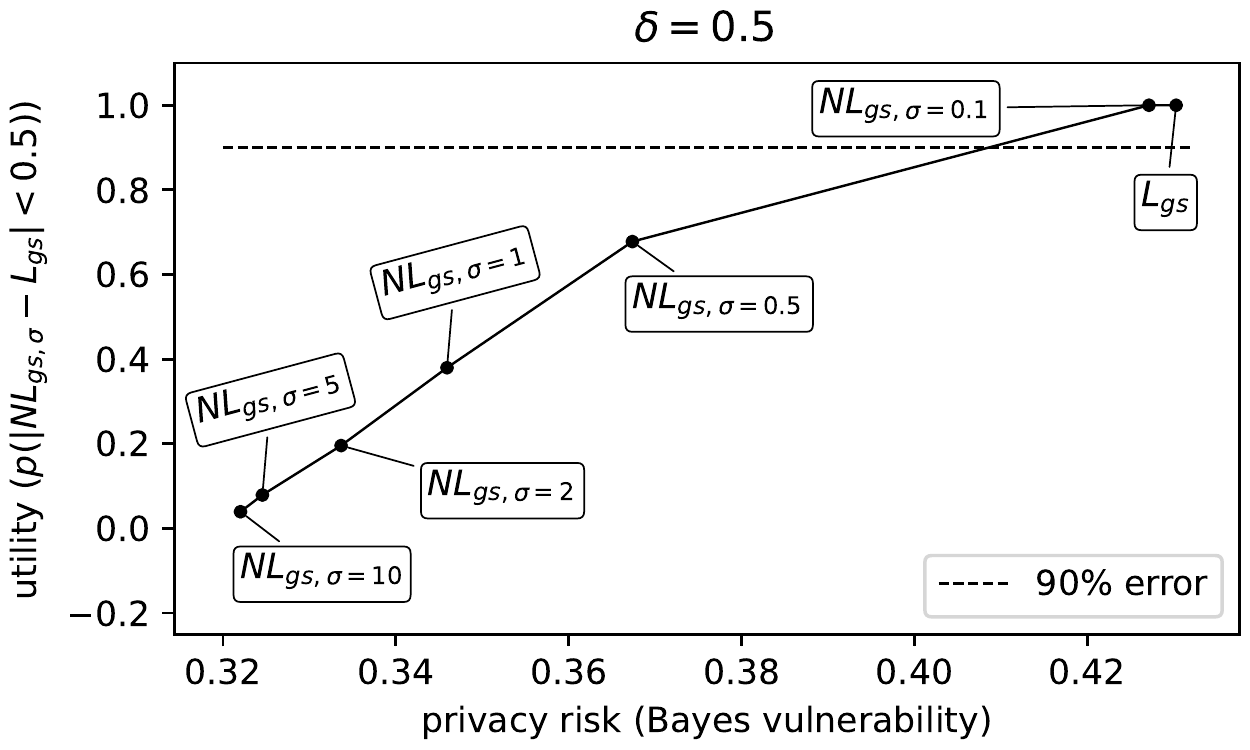}
  \includegraphics[trim=50 0 0 0, clip, width=.31\textwidth]{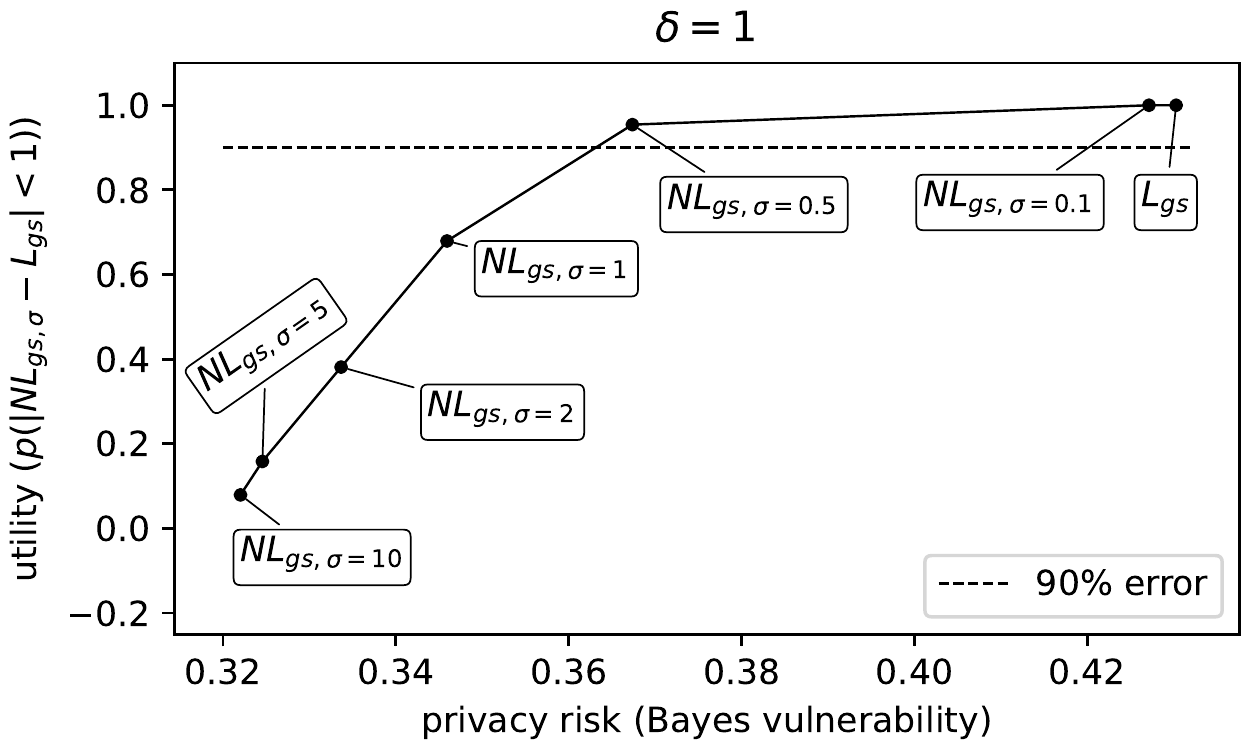}
  \caption{Utility/privacy trade-off for all polygenic scores for different error bounds.}
  \label{fig:results-utility-bayes-vulnerability}
\end{figure*}

\begin{figure*}[t!]
  \centering
  \includegraphics[width=\textwidth]{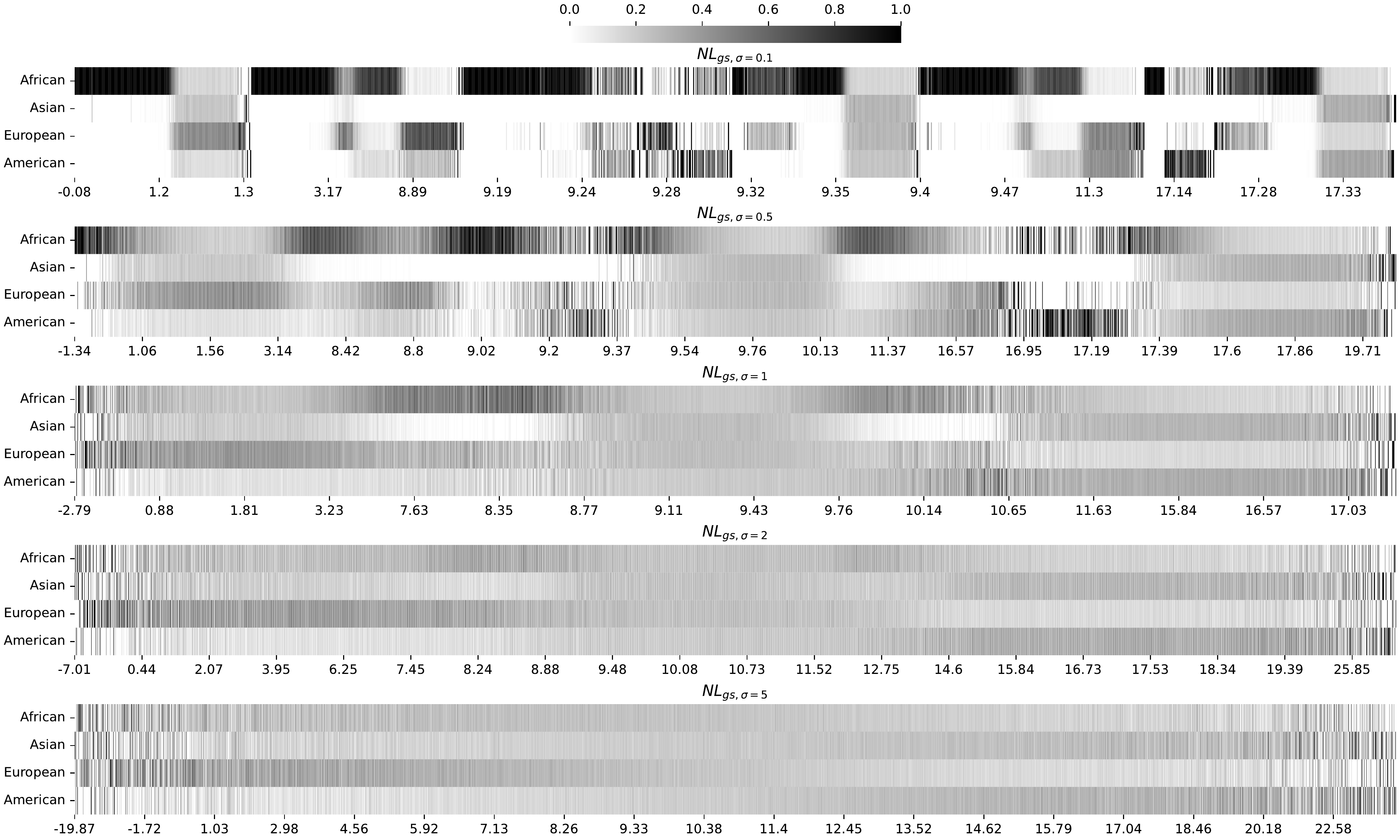}
  \includegraphics[width=\textwidth]{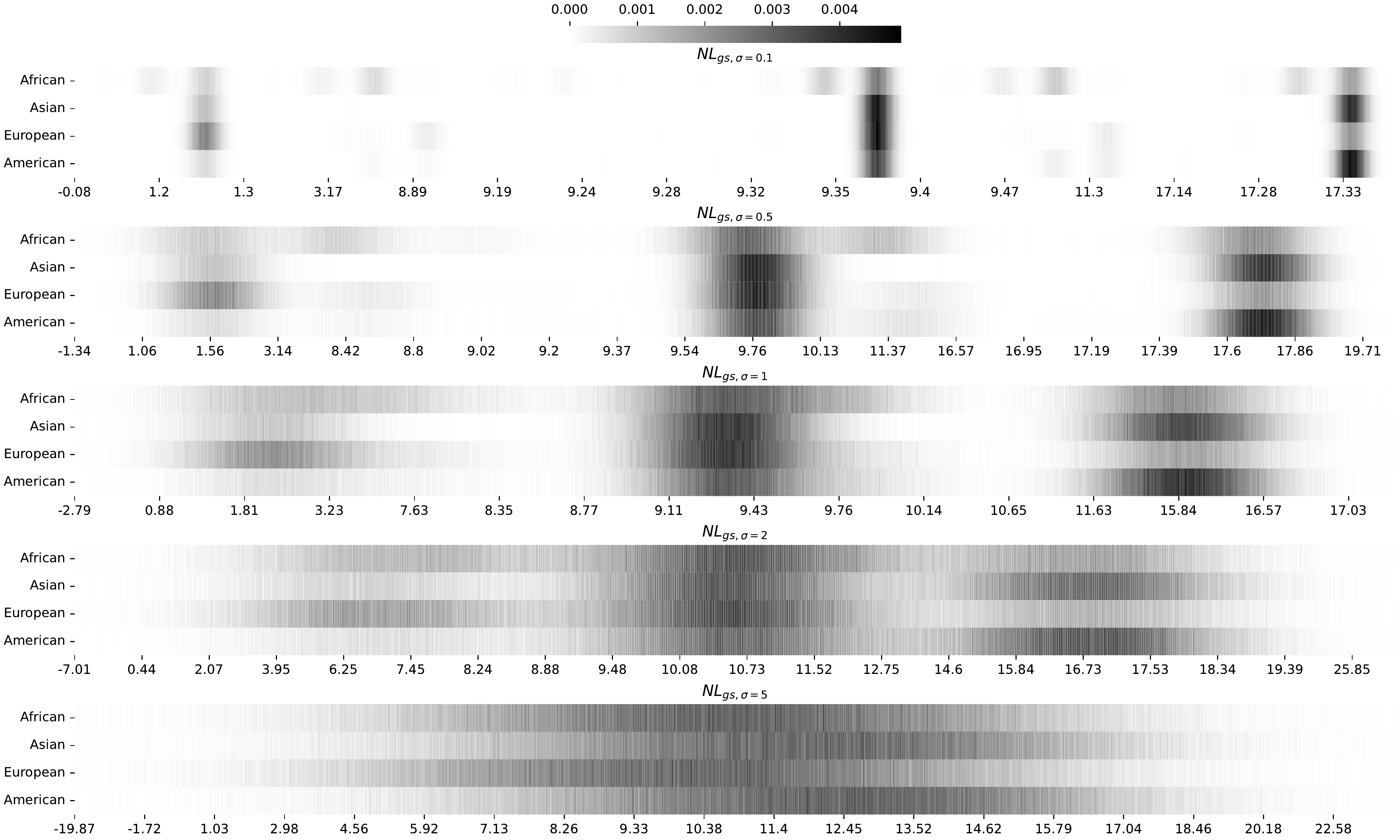}
  \caption{
    Output privacy risk (top) and program privacy risks (bottom) for polygenic scores with random noise for increasing $\sigma$.
  }
  \label{fig:results-noisy-linear-gs-conditional}
  \label{fig:results-noisy-linear-gs-joint}
\end{figure*}


\subsection{Posterior inference}
\label{subsec:posterior-inference}
To estimate the joint posterior distribution, we use a Metropolis-within-Gibbs sampler optimized for Categorical variables~\cite{handbook_mcmc,pymc3}.
%
The model in~\cref{sec:modelling} is composed by ethnicity and haplotype variables which are in a nominal (categorical) scale.
We generate $100$k samples with a burn-in period of $50$k samples.
The resulting posterior distribution shows good sampling/convergence diagnosis~\cite{handbook_mcmc}: Estimated Sample Size (ESS) of at least $50$k, a Markov Chain Standard Error (MCSE) below $0.15$, and $\hat{R}$ of $1.0$ for all parameters.
This diagnosis indicates that the inferred posterior has converged and it is accurate.

\subsection{Output Privacy Risk}
\label{sec:results-output-privacy}

\noindent
\textit{Binary tasting phenotype.}
\label{subsub:results-output-privacy-phenotype}
The last two heatmaps in~\cref{fig:results-phenotype} (top-left) show the output privacy results for the tasting phenotype program, for TAS2R38 and TAS2R16, respectively.
We observe that the non-taster output carries higher privacy risks in both cases.
For TAS2R16, it implies completely giving away the ethnicity of the individual.
Interestingly, the taster output is (mostly) uniformly distributed among ethnicities.
This means that in both cases it is safe to publish that the individual is a taster.

\noindent
\textit{Wine tasting score.}
\label{subsub:results-output-privacy-linear-score}
\Cref{fig:results-linear-gs-conditional} (middle) shows the output privacy risk of the wine tasting score.
For more than half of the possible outputs, the African ethnicity is at high risk, \textit{i.e.}, output privacy risk close to 1.
European is the second most vulnerable ethnicity.
The American ethnicity shows high  risk only for 3 possible outputs.
The Asian ethnicity shows low risk for all outputs.
For the outputs $9.31$ and $17.37$ the output privacy risk of each ethnicity is very close to $0.25$ (the same as in the prior).
This means that the attacker would not learn much by observing this output.
For data analysts interested in output privacy risk, we recommend to only disclose the output if it equals $9.31$ or $17.31$.

\noindent
\textit{Wine tasting score with random noise.}
\label{subsub:results-output-privacy-noisy-linear-score}
\Cref{fig:results-noisy-linear-gs-conditional} (top) shows the output privacy results of the wine tasting score with random noise for an increasing noise level $\sigma$ from $0.1$ to $5$.

%
For $\sigma=0.1$, Africans and Europeans have higher output privacy risk.
%
%
Due to the large number of outputs, now we discuss ranges of possible outputs.
A gray homogeneous color indicates that risk is distributed uniformly across ethnicities.
We observe this effect in the range $(9.35, 9.40)$ and around $17.33$.
These intervals are close to the low risk values in the wine tasting score without noise.
Values of $\sigma$ equal to $0.5$ and $1$ increase the width of the uniformly distributed areas.
This effect also reduces the size of solid black ranges, meaning that output privacy improves, especially for Africans and Europeans.
For values of $\sigma$ greater than $1$ the above effect is more pronounced.
As $\sigma$ increases, a large uniform gray range (with low output privacy risks) covers most of the spectrum of output values.
For $\sigma=2$ and $\sigma=5$, outputs (approximately) in the range $(9.5, 11.5)$ show good output privacy.
We also observe that, in these cases, the border regions (low and high output values) show high contrast indicating high output privacy risks for some ethnicities.

Our results indicate that for $0 < \sigma < 1$ outputs in the regions close to $9.31$ and $17.37$ have low output privacy risk.
For $1 < \sigma \leq 5$, outputs in the range $(9.5,11.5)$ show low output privacy risks.
As expected, the larger $\sigma$ the higher the privacy, but we compromise its utility. We discuss this in~\cref{sec:results-utility}.

\subsection{Program Privacy Risk}
\label{sec:results-program-privacy}

\noindent
\textit{Binary tasting phenotype.}
\label{subsub:results-program-privacy-phenotype}
%
%
For TAS2R16 (\cref{fig:results-phenotype}, 2nd heatmap in top-left), we observe that \emph{taster} has the highest probability and is uniformly distributed across ethnicities.
This indicates (almost) complete absence of program privacy risks.
Recall that \emph{non-taster} has very high output privacy (\cref{subsub:results-output-privacy-phenotype}).
Now program privacy reveals that the non-taster output has very low probability; only $0.01$. (Probability values do not add up to 1 because they are rounded.)
TAS2R38 (\cref{fig:results-phenotype}, 1st heatmap in top-left) shows similar results:
Taster is a more likely output than non-taster, and probability is distributed uniformly across ethnicities for taster.
We observe that for tasters, Europeans have slightly lower program privacy risk than the other ethnicities.
However, when the output is non-taster, Europeans have double the program privacy risks compared to other ethnicities; with this scenario occurring with probability 0.06.
This is inline with the high output privacy risks for Europeans and this program output, but program privacy shows that this case is unlikely.

To sum up, both binary tasting genotype programs have low program privacy risks, with TAS2R16 offering better protecton than TAS2R38.

\noindent
\textit{Wine tasting score.}
\label{subsub:results-program-privacy-linear-score}
\Cref{fig:results-linear-gs-joint} (bottom) shows the program privacy risks for the wine tasting score.
We observe that only outputs 1.24, 9.31 and 17.37 have non-negligible probability.
This is useful information, as output privacy allocated high privacy risks for Africans, but now we discover that those outputs are very unlikely.
In fact, within the high probability outputs, Africans show the lowest risk of all ethnicities.
Interestingly, the outputs with non-negligible probability coincide with some of those having low output privacy risks, \textit{i.e.,} 9.31 and 17.37.
As for output 1.24, although it does not exhibit high program privacy risks, the probability for Europeans is higher than for others.

All in all, the wine tasting score shows a good level of program privacy risks.
Probability is mostly distributed across ethnicities for all likely outputs.
However, this distribution is less uniform than for the binary tasting phenotype.
This is expected as the polygenic score contains genetic information about TAS2R38 \emph{and} TAS2R16.
On the contrary, the phenotype programs work on either TAS2R38 or TAS2R16.

\noindent
\textit{Wine tasting score with noise.}
\label{subsub:results-program-rivacy-noisy-linear-score}
\Cref{fig:results-noisy-linear-gs-joint} (bottom) shows the results of program privacy risks for the wine tasting score with random noise.
Each row displays the results for a value of $\sigma$, starting from $\sigma=0.1$ up to $\sigma=5$.

In the first 3 rows ($0.1 \leq \sigma \leq 1$), we observe 3 distinct high probability regions.
Note that these coincide with the high probability outputs in the program privacy risks for the wine tasting score without noise: 1.24, 9.31 and 17.37.
Similarly to the score w/o noise, program privacy reveals that most of the high risk outputs for Africans are unlikely events.
Also, program privacy is (mostly) uniformly distributed across ethnicities for the high probability outputs.
With higher program privacy risk for:
\begin{inparaenum}[i)]
\item Europeans in outputs around 1.24, Asian and Europeans;
\item Europeans and Asians in outputs around 9.31; and
\item Asians and Americans for outputs around 17.37.
\end{inparaenum}
Nevertheless, these results are positive, as there is no ethnicity with significantly higher risks.
For the last 2 rows ($2 \leq \sigma \leq 5$), the 3 regions above merge into a single high probability region centered around 10.
Program privacy risks across ethnicities become more uniform as $\sigma$ grows.
This is displayed as a uniform gray tone across ethnicities.
This is a clear indication of low program privacy risks.
As mentioned earlier, these results must be considered together with utility metrics.
We discuss utility in~\cref{sec:results-utility}.

To sum up, random noise improves the program privacy of the wine tasting score,
especially for large values of $\sigma$ where privacy risks are uniformly distributed across ethnicities.
It is unclear, however, how it compares with the binary phenotype programs.
The following section, that discusses the results of program privacy scores, will allow us to effectively compare all disclosure programs.

\subsection{Privacy Risk Scores}
\label{sec:results-program-privacy-scores}

\noindent
\textit{Maximum output privacy.}
\label{subsub:results-maximum-output-privacy}
\Cref{fig:results-max-output-privacy-risk} (top-right, light gray columns) shows the maximum output privacy risk for all disclosure programs.
We observe that all programs except for $\Ph^\rthirtyeigth$ have max.\ output privacy risk 1.
This is because they have at least one output for which output privacy risk equals 1.
As discussed earlier, this metric is quite pessimistic: we saw in the program privacy results that most of the outputs with high output privacy are very unlikely.
Nevertheless, maximum output privacy serves as a good upper bound on risk.
Here we can see that $\Ph^\rthirtyeigth$ is close to 0.4.
This means that, no matter the output, output privacy risks will never be above this value.
This may be a sufficient level of privacy, taking into account that 0.25 is the prior probability for each ethnicity.

\noindent
\textit{Bayes vulnerability.}
\label{subsub:results-bayes-vulnerability}
\Cref{fig:results-bayes-vulnerability} (top-right, dark gray columns) shows the Bayes vulnerability results for all disclosure programs.
Recall that Bayes vulnerability measures the expected probability of learning ethnicity by observing the output.
This metric scales the risk in each output by the probability of the output.
As a consequence, we observe lower risk levels when compared with maximum output privacy.
Interestingly, $\Ph^\rsixteen$ has a lower risk score than $\Ph^\rthirtyeigth$ (as opposed to what we observed in maximum output privacy).
This is because the output with high output privacy risk in $\Ph^\rsixteen$ is very unlikely.
As expected, the wine tasting score without noise has the largest Bayes vulnerability; as it encapsulates the most information.
The results show that the effect of noise reduces the risk, but not substantially.
None of the levels of noise we analyzed show lower privacy risks than the binary tasting phenotype programs.
However, the maximum Bayes vulnerability is $\approx 0.42$, which is not a very high value.

There is no universal value for perfect Bayes vulnerability.
Companies/institutions may fix values for Bayes vulnerabilities based on their privacy requirements.
For illustrative purposes, we (arbitrarily) set on a value no more than 0.35 Bayes vulnerability, \textit{i.e.}, at most 0.1 more than the prior.
Then, only the tasting phenotype programs and wine tasting score with random noise and $\sigma \geq 0.5$ are considered privacy preserving.

\subsection{Utility}
\label{sec:results-utility}

\noindent
\textit{Absolute distance distribution.}
\label{subsub:results-utility-absolute-distance}
\Cref{fig:results-utility-metric} (left,center) shows the absolute distance distribution between the wine tasting score with and without noise for different values of $\sigma$.
The results are split into two figures to better appreciate the High Density Interval (HDI) of the distributions; note the difference range values for x and y axes.
For $0.1 \leq \sigma \leq 1$ (left plot), all distributions have their mode close to 0 and the HDI ends around $1$ (or $1.5$ for $\sigma=1$).
This indicates a small introduced error.
For the right plot ($\sigma > 1$), the mode is close to $0$ as well, but the HDIs are much wider:
The HDI ends at $5$ for $\sigma=2$, at $10$ for $\sigma=5$ and at $15$ for $\sigma=10$.

The extent to which the error in the linear score is admissible is problem dependent.
However, this analysis shows the large amount of distortion that values of $\sigma > 0.5$ introduce.
Fixing a maximum level of error would be helpful in deciding what programs have acceptable error.
The next metric explores this.

\noindent
\textit{Error bound probability.}
\label{subsub:results-utility-error-bound-probability}
\Cref{fig:results-utility-metric} (right) shows the error bound probability results for increasing value of $\sigma$ in the wine tasting score program with noise for error bound values $\delta \in \{0.1, 0.5, 1\}$.
We consider these values acceptable given the scale of the wine tasting score.
However, the value of $\delta$ is application dependent, and our method can be used with any value of $\delta$.
%
%
%
The plots show the 90\% probability boundary, which we consider sufficient confidence. Stronger requirements can be set, e.g., 95\% or even 100\%.

For all $\delta$ values, we observe a sharp exponential decay in utility as $\sigma$ increases.
For $\delta=0.1$, utility decays below 20\% even for $\sigma=0.5$.
For $\sigma > 0.1$, we observe worse utility: with values very close to 0.
This indicates that \emph{only} values of $\sigma$ close to 0.1 are acceptable.
Increasing the error bound to $\delta=0.5$ yields better utility.
Yet no value $\sigma > 0.1$ meets our requirements.
It is only for $\delta=1$ that $\sigma=0.5$ meets our utility requirements.
As for the other cases, no value $\sigma > 0.5$ has acceptable utility.

Adding noise decreases privacy, but it is not a panacea.
The results in this section show that we can only add a small amount of noise, if we want to preserve utility.

\noindent
\textit{Privacy/Utility trade-off.}
\label{subsec:results-privacy-utility-trade-off}
We conclude by putting together the privacy and utility scores.
\Cref{fig:results-utility-bayes-vulnerability} plots the error bound probability (utility) and Bayes vulnerability (privacy).
We analyze different levels of $\delta$, as before.

For $\delta=0.1$ (left in~\cref{fig:results-utility-bayes-vulnerability}), we observe that no level of noise meets the utility requirements.
Only $\Lgs$ (wine tasting score w/o noise) is above the 90\% line.
$\NLgs$ with $\sigma=0.5$ shows a utility level around 60\% with almost the same Bayes vulnerability.
In other words, we gain no privacy and deteriorate utility to an unacceptable degree (for $\delta=0.1$).
Reducing the utility requirements to $\delta=0.5$ (center in~\cref{fig:results-utility-bayes-vulnerability}) includes $\sigma=0.1$ as an acceptable program.
But, again, we gain almost no privacy protection.
Finally, for $\delta=1$, $\NLgs$ with $\sigma=0.5$ meets the utility requirements.
In this case, Bayes vulnerability is reduced by 0.06 (from 0.43 to 0.37).
This privacy score is still far from the prior (\textit{i.e.,} 0.25), but it is a significant improvement.

Data analysts may use these results to make an informed decision on the programs to disclose the wine tasting score.
Given our results and privacy/utility requirements we set forth, the best choice would be $\NLgs$ with $\sigma=0.5$.
That said, the most valuable takeaway is the analysis process and privacy/utility information we described.











%% file: related.tex
\section{Related Work}
\label{sec:related}

There exists a wide spectrum of research on genetic privacy~\cite{acm_survey_genomic_privacy}. 
Below we cover the most relevant work in the context of this paper.

%
Cai~\etal~\cite{2015_cai} develop a re-identification attack based on Genome-Wide Association Studies (GWAS).
These studies are applied to human genomic data to understand disease associations.
The presented algorithm scales well for realistic GWAS datasets.
They show that the number of re-identified individuals grows with number of released genotypes.
Gymrek~\etal~\cite{2013_gymrek} demonstrate re-identification risks by combining haplotype information with demographics such as age and state.
In particular, they analyze the probability of re-identifying US males.
Our work focuses on privacy risks associated to phenotypes and polygenic scores instead of working directly on genotypes.
Also, we focus on the problem of inferring sensitive data (ethinicity) as opposed to re-identification.

%
G{\"u}rsoy~\etal~\cite{2020_Gursoy_and_Gerstein} study the probability of inferring \emph{sensitive} phenotypes, \textit{i.e.}, phenotypes the victim wants to keep secret.
The consider an attacker with access to public studies on correlation between genotypes and sensitive phenotypes.
Given the genotype of a victim, they compute the probability of learning the sensitive phenotypes.
The authors propose a data sanitation protocol for genotypes that minimizes the probability of learning sensitive phenotypes.
Similarly, Harmancie and Gerstein~\cite{2018_harmanci} study privacy risks on genomic deletions on signal profiles.
%
%
Genomic deletions may enable attacker to infer sensitive phenotypes via public statistics on the correlation between deletions and phenotypes.
The authors propose an anonymization method based on removing dips in signal profiles.
These works tackle the problem of inferring sensitive phenotypes from genotype data.
Instead, we quantify and protect against inferring sensitive data from public polygenic scores or phenotypes.
\looseness = -1

%
Humbert~\etal~\cite{humbertCCS13,humbertTops2017} propose a probabilistic model to infer Single Nucleotide Polymorphism (SNP) values. 
%
%
They use an inference algorithm (belief propagation or Bayesian inference) to  estimate the distribution of unknown SNP values from information about observed SNPs, genomic data of family members, familial relationships, etc.
The authors also define health privacy scores based on SNP values. 
Anonymization is performed by masking specific SNPs. 
They further propose an optimization algorithm that determines the SNP to mask to minimize risks in the aforementioned model~\cite{humbertWPES2014}.
In this context, Humbert~\etal~\cite{humbertEuroSP2022} have developed a tool for communicating and raising aware of kin privacy to lay users.
These works focus on SNP information to quantify privacy risks, we instead target polygenic scores and phenotypes.

%% file: conclusion.tex
\section{Conclusion}
\label{sec:conclusion}

Polygenic risk scores are typically defined as a weighted sum on genetic data related to a single phenotype trait. They are used to summarize the effect of genes on phenotypes, both to inform the individual patient and to anonymize results for publication. As discussed above, any disclosure of genetic data including polygenic scores is associated with the risk of re-identifying individuals or to find out a predisposition to a disease.

In this paper, we have introduced an approach to quantify and prevent privacy risks by focusing on polygenic scores and phenotypic information.
We believe that this is the first work to explore this viewpoint to tackle genetic privacy.
Building on top of the privacy risk analysis method \privug~\cite{privug}, 
we compute the attacker posterior knowledge from a program to compute the polygenic risk score, a probabilistic model of attacker knowledge about the individuals covered and populations, and an output of the program. 

Our approach aims at supporting existing methods with a novel way to measure the risk of privacy violations.
We have demonstrated its application on a polygenic trait score for the TAS2R38 and TAS2R16 taste receptor genes. We have shown how to quantify the risks for a person's privacy in regards to their ancestry and thereby derived their likely ethnicity.
While the data and programs in this case study were selected to enhance readability, our methodology can be applied to phenotypes and polygenic scores working on any kind of sensitive genetic data.

%% file: main.bbl
\begin{thebibliography}{10}
\providecommand{\url}[1]{\texttt{#1}}
\providecommand{\urlprefix}{URL }
\providecommand{\doi}[1]{https://doi.org/#1}

\bibitem{qif}
Alvim, M.S., et~al.: The Science of Quantitative Information Flow. Information
  Security and Cryptography, Springer (2020)

\bibitem{2013_Behrens}
Behrens, M., et~al.: Genetic, functional, and phenotypic diversity in
  {TAS2R38}-mediated bitter taste perception. Chem. Senses  \textbf{38},
  475--484 (2013)

\bibitem{handbook_mcmc}
Brooks, S., Gelman, A., Jones, G., Meng, X.L.: Handbook of markov chain monte
  carlo. CRC (2011)

\bibitem{2002_Bufe}
Bufe, B., et~al.: The human {TAS2R16} receptor mediates bitter taste in
  response to $\beta$-glucopyranosides. Nature Genetics  \textbf{32}(3),
  397--401 (2002)

\bibitem{2015_cai}
Cai, R., Hao, Z., Winslett, M., Xiao, X., Yang, Y., Zhang, Z., Zhou, S.:
  Deterministic identification of specific individuals from {GWAS} results.
  Bioinform.  \textbf{31}(11),  1701--1707 (2015)

\bibitem{wine_tasting}
Carrai, M., et~al.: Association between taste receptor ({TAS}) genes and the
  perception of wine characteristics. Scientific Reports  \textbf{7}(1) (2017)

\bibitem{2013_gymrek}
Gymrek, M., McGuire, A.L., Golan, D., Halperin, E., Erlich, Y.: Identifying
  personal genomes by surname inference. Science  \textbf{339}(6117),  321--324
  (2013)

\bibitem{2020_Gursoy_and_Gerstein}
Gürsoy, G., et~al.: Data sanitization to reduce private information leakage
  from functional genomics. Cell  \textbf{183},  905--917 (2020)

\bibitem{2018_harmanci}
Harmanci, A., Gerstein, M.: Analysis of sensitive information leakage in
  functional genomics signal profiles through genomic deletions. Nature
  Communications  \textbf{9}(1) (2018)

\bibitem{numpy}
Harris, C.R., et~al.: Array programming with {NumPy}. Nature
  \textbf{585}(7825),  357--362 (2020)

\bibitem{humbertCCS13}
Humbert, M., Ayday, E., Hubaux, J., Telenti, A.: Addressing the concerns of the
  lacks family: quantification of kin genomic privacy. In: CCS. pp. 1141--1152.
  {ACM} (2013)

\bibitem{humbertWPES2014}
Humbert, M., Ayday, E., Hubaux, J., Telenti, A.: Reconciling utility with
  privacy in genomics. In: Workshop on Privacy in the Electronic Society,
  {WPES}. pp. 11--20. {ACM} (2014)

\bibitem{humbertTops2017}
Humbert, M., Ayday, E., Hubaux, J., Telenti, A.: Quantifying interdependent
  risks in genomic privacy. {ACM} Trans. Priv. Secur.  \textbf{20}(1),
  3:1--3:31 (2017)

\bibitem{humbertEuroSP2022}
Humbert, M., Didier, D., Mauro, C., K\'evin, H.: {KGP} meter: Communicating kin
  genomic privacy to the masses. In: {EuroS\&P}. {IEEE} (2022)

\bibitem{2020_Lumsden}
Lumsden, A.L., et~al.: Apolipoprotein {E} ({APOE}) genotype-associated disease
  risks: a phenome-wide, registry-based, case-control study utilising the {UK}
  biobank. EBioMedicine  \textbf{59}(102549) (2020)

\bibitem{acm_survey_genomic_privacy}
Naveed, M., Ayday, E., Clayton, E.W., Fellay, J., Gunter, C.A., Hubaux, J.,
  Malin, B.A., Wang, X.: Privacy in the genomic era. {ACM} Comput. Surv.
  \textbf{48}(1),  6:1--6:44 (2015)

\bibitem{privug}
Pardo, R., et~al.: Privug: Using probabilistic programming for quantifying
  leakage in privacy risk analysis. In: ESORICS. LNCS, vol. 12973. Springer
  (2021)

\bibitem{2016_Risso}
Risso, D.S., et~al.: Global diversity in the {TAS2R38} bitter taste receptor:
  revisiting a classic evolutionary proposal. Scientific Reports  \textbf{6}(1)
  (2016)

\bibitem{pymc3}
Salvatier, J., Wiecki, T.V., Fonnesbeck, C.: Probabilistic programming in
  {Python} using {PyMC3}. PeerJ Comput. Sci.  \textbf{2}, ~e55 (2016)

\bibitem{SORANZO20051257}
Soranzo, N., et~al.: Positive selection on a high-sensitivity allele of the
  human bitter-taste receptor {TAS2R16}. Current Biology  \textbf{15}(14),
  1257--1265 (2005)

\bibitem{2018_polygenetic_risk_scores}
Torkamani, A., Wineinger, N.E., Topol, E.J.: The personal and clinical utility
  of polygenic risk scores. Nature Reviews Genetics  \textbf{19}(9),  581--590
  (2018)

\end{thebibliography}
